\documentclass{PoS}
\newcommand{\be}{\begin{equation}}
\newcommand{\ee}{\end{equation}}
\newcommand{\bdm}{\begin{displaymath}}
\newcommand{\edm}{\end{displaymath}}
\newcommand{\bea}{\begin{eqnarray}}
\newcommand{\eea}{\end{eqnarray}}

\newcommand{\fs}{\; .}
\newcommand{\co}{\; ,}

\newcommand{\lbar}{\bar{\ell}}

\newcommand{\LO}{\,=\hspace{-0.37cm}\rule[0.35cm]{0cm}{0cm}^{\mbox{\tiny LO}}\hspace{0.15cm}}
\newcommand{\NLO}{\,=\hspace{-0.44cm}\rule[0.35cm]{0cm}{0cm}^{\mbox{\tiny NLO}}\hspace{0.08cm}}
\newcommand{\p}{\raisebox{0.05cm}{\tiny $\bullet$}\hspace{0.2cm}}
\title{Light quark masses}

\ShortTitle{Light quark masses}

\author{\speaker{H.~Leutwyler} \\
       Albert Einstein Center for Fundamental Physics\\
Institute for Theoretical Physics, University of Bern\\ Sidlerstrasse 5, 3012 Bern, Switzerland\\
       E-mail: \email{Leutwyler@itp.unibe.ch}}

\abstract{Low energy precision experiments provide significant tests of the laws of nature that can reveal physics beyond the Standard Model. A good theoretical understanding of the low energy properties of QCD is required for this purpose. The recent developments at the interface between lattice and effective field theory methods provide an excellent basis for pion physics already now, while the extension required to explore the low energy properties in the strange quark sector yet calls for further work, also in view of a better determination of the light quark masses. 

At the precision achieved in lattice determination of quark mass ratios, the  e.m.~self-energies of the mesons play an important role. I point out some unresolved issues occurring in this context and then summarize the present knowledge of the light quark masses.
}

\FullConference{6th International Workshop on Chiral Dynamics\\
		 July 6-10 2009\\
		 Bern, Switzerland}

\begin{document}

\section{Effective low energy theory for QCD}
At low energies ($E\ll M_W$), the weak interactions are frozen, so that the Standard Model reduces to ${\cal L}_{QCD} + {\cal L}_{QED}$. This Lagrangian specifies a comprehensive precision theory for cold matter ($T\ll M_W$), which determines the size of the atoms, the structure of solids, the nuclear reactions in the sun, etc. Moreover, ${\cal L}_{QED}$ is infrared stable and is characterized by a pure number that happens to be small. The effects due to the electromagnetic interaction can therefore be accounted for with perturbation theory. At low energies and over distances that are not too large, so that gravitational effects are weak, the known laws of nature reduce to QCD + corrections. The pi\` ece de r\'esistance in this domain is Quantum Chromodynamics.  

There are many  models that resemble QCD in one way or the other: instantons, monopoles, bags, superconductivity, gluonic strings, linear $\sigma$ model, NJL, hidden gauge, AdS/CFT, \ldots\,, but none of these has led to an approximation scheme that would allow us, at least in principle, to solve the theory. The observation confirms the experience that miracles are rare, particularly in physics.  

Since nonperturbative methods are required to analyze the low energy properties of QCD in a model independent way, the development in this field has been slow. Nevertheless, considerable progress has been achieved in recent years, with numerical simulations on a lattice, with sum rules and dispersion relations and with effective field theory, which in this context is referred to as Chiral Perturbation Theory ($\chi$PT). In all of these methods, {\it chiral symmetry} plays an essential role: the fact that the Hamiltonian of QCD with $N_f$ massless quarks has an exact SU$_L(N_f)\times$SU$_R(N_f)$ symmetry. We also know that this symmetry is "hidden" or "spontaneously broken", the ground state being invariant only under the subgroup SU$_{L+R}(N_f)$. 

The symmetry group SU$_L(N_f)\times$SU$_R(N_f)$ is broken not only spontaneously, but also explicitly, by the quark mass term in the Lagrangian of QCD. It so happens, however, that the two lightest quarks are very light. Hence the part of the Lagrangian that breaks the subgroup SU$_L(2)\times$SU$_R(2)$ -- the mass term of the $u$- and $d$- quarks -- is very  small:  the theory does not have an exact symmetry, but an approximate one that is nearly exact. In fact, already in 1960, Nambu \cite{Nambu} found out that (i) the strong interaction must have an approximate SU(2)$\times$SU(2) symmetry, (ii) for dynamical reasons, the ground state is invariant only under the isospin subgroup SU(2), so that the symmetry breaks down spontaneously, (iii) the spontaneous breakdown of an exact Lie group symmetry entails massless particles (iv) for the strong interaction, the pions play this role and (v) the pions are not massless, only light, because the symmetry is only an approximate one. The occurrence of {\it approximate} symmetries in nature, which at that time looked mysterious, has found a natural  explanation within QCD: it so happens that $m_u$ and $m_d$ are very small.
\begin{figure}[thb]\label{fig:a0a2}
\centering
\includegraphics[width=8.3cm]{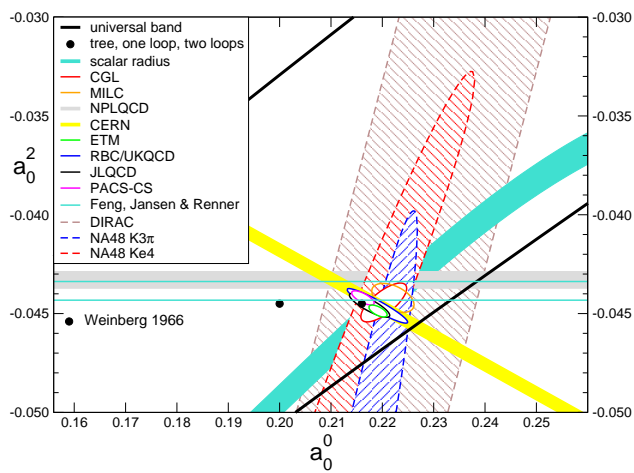}
\caption{Experimental and theoretical results for the S-wave $\pi\pi$ scattering lengths}
\end{figure}

In the meantime, it has been shown that the low energy properties of QCD can be analyzed quantitatively and in a model independent way by treating $m_u$ and $m_d$ as perturbations. On this basis, pion physics has become a precision laboratory. Low energy precision experiments on the magnetic moment of the muon, on kaon decays and $\pi^+\pi^-$ atoms provide very significant tests of the theory that can reveal physics beyond the Standard Model. As an illustration, Figure \ref{fig:a0a2} compares the experimental and theoretical results for the $\pi\pi$ S-wave scattering lengths and shows that the theoretical predictions are confirmed to remarkable accuracy. Some of the entries shown in this plot were discussed in detail at the workshop and are described in contributions to these proceedings. The ellipse marked CGL \cite{CGL} is obtained by matching the NNLO representation of $\chi$PT with the dispersive representation of the scattering amplitude below threshold, where those higher order corrections that concern the momentum dependence are particularly small. Most of the lattice results shown are obtained by measuring the dependence of $M_\pi$ and $F_\pi$ on $m_u,m_d$, using the data to determine the coupling constants $\lbar_3$, $\lbar_4$ that occur in the effective Lagrangian and inserting the result in the $\chi$PT formulae for $a_0^0$, $a_0^2$ -- the errors are dominated by those in $\lbar_3$, $\lbar_4$ (for a recent discussion of the uncertainties due to the NNLO coupling constants, I refer to \cite{Guo:2009bt}). At the accuracy reached in this field, it is important that the e.m.~interaction is properly account for \cite{em corrections}.  

I add a remark concerning pion form factors, which can now also be calculated on the lattice. The {\it electromagnetic pion form factor} is known very well from dispersion theory: the experimental information obtained from $e^+e^-\rightarrow \pi^+\pi^-$ and $\pi^\pm\, e\rightarrow \pi^\pm\, e$ suffices to reliably evaluate the relevant dispersion integrals and to determine the form factor very accurately. The lattice results do not yet reach the precision to which the e.m.~form factor is known, but the calculations offer an excellent testing ground for the lattice method. The {\it scalar pion form factor} is of particular interest, because it reflects aspects of the low energy structure of QCD that are crucial for our understanding, but are not accessible to experiment -- nature is not kind enough to offer us a sufficiently light scalar boson that could be used as a probe analogous to the photon. At NLO, the low energy behaviour of the scalar form factor is determined by the same two low energy constants $\lbar_3,\lbar_4$ that were discussed above. The value at $t=0$ is connected with the $\sigma$-term of the pion, i.e.~with the logarithmic derivative of $M_\pi^2$ with respect to $m_u,m_d$. First results for the scalar radius \cite{Aoki proceedings} are consistent with the theoretical prediction \cite{CGL}, but the uncertainty is yet too large to draw conclusions. The lattice determination of the scalar form factor is of great interest and should vigorously be pursued.

\section{Expansions in the mass of the strange quark}
In the limit where all three light quarks are taken massless, QCD acquires an exact symmetry under SU(3)$_L\times$SU(3)$_R$. Chiral Perturbation Theory can be extended accordingly, treating not only $m_u,m_d$, but also $m_s$ as a perturbation. The key question in this extension is whether the physical value of $m_s$ is small enough for  SU(3)$_L\times$SU(3)$_R$ to represent a useful approximate symmetry. The conviction that this must be the case originates in the success of the eightfold way of Gell-Mann and Ne'eman: there is good evidence that the group SU(3)$_{L+R}$ does represent a decent approximate symmetry (multiplet pattern, approximate validity of the Gell-Mann-Okubo formulae, prediction of the $\Omega^-$ baryon, approximate equality of the decay constants of pion and kaon, etc.). In the framework of QCD, the only plausible explanation of this observation that I am aware of is that the quark mass differences $m_s-m_d$, $m_d-m_u$, which are responsible for the level splittings within the multiplets, are small and can be treated as perturbations. The fact that $M_K^2$ is much larger than $M_\pi^2$ implies that $m_s$ is much larger than $m_u$ or $m_d$. Hence $m_s$ itself must represent a small perturbation, so that SU(3)$_L\times$SU(3)$_R$ must be an approximate symmetry as well. Since $m_s-m_d\simeq m_s$, the symmetry breaking is of the same size as the one for SU(3)$_{R+L}$. The fact that the squares of the Nambu-Goldstone masses obey the Gell-Mann-Okubo formula remarkably well indicates that the tree level formulae for these masses, which follow from the extension of $\chi$PT to SU(3)$_L\times$SU(3)$_R$, do represent a good approximation. I conclude that the expansion in powers of $m_u,m_d,m_s$ ought to work, but a comparatively slow convergence is to be anticipated.

The terms occurring in the chiral perturbation series of the masses and decay constants can now be calculated on the lattice.\footnote{In the present report, there is not enough space for a comparison of the results obtained by the various lattice collaborations. I restrict myself to the MILC data. For a comprehensive discussion of the lattice results relevant for low energy particle physics, I refer to the forthcoming review by the FLAG working group  of Flavianet \cite{FLAG}.}  In accordance with the comments given in the preceding section, the $\chi$PT predictions for the dependence of the pion decay constant on $m_u,m_d$ are confirmed. The expansion in powers of these masses is dominated by the leading term, which represents the value of $F_\pi$ in the limit $m_u,m_d\rightarrow 0$ and is denoted by $F$. The lattice result, $F_\pi/F= 1.062(1)(3)$ \cite{Bazavov:2009fk},  is consistent with the prediction $F_\pi/F=1.072(7)$ \cite{Colangelo and Duerr}. 

If the OZI rule were exact, $F_\pi$ would be independent of $m_s$. This suggests that $F$ is close to the limiting value $F_0$, obtained if all three light quark masses are set equal to zero. The lattice result, $F/F_0=1.10(4)$ \cite{Bazavov:2009fk}, indicates that the violations of the OZI rule are indeed of modest size. If the expansion in powers of $m_s$ is truncated at NLO, the $\chi$PT formula for the ratio $F/F_0$ only involves the effective coupling constant $L_4$. With the value obtained by MILC, $\chi$PT yields $F/F_0\NLO 1.09(3)$. This shows that, in the expansion of $F/F_0$ in powers of $m_s$, the corrections are dominated by the term proportional to $m_s$ -- those of higher order are hidden in the noise of the calculation. 

The result $F_K/F_\pi=1.198(2)(^{+6}_{-8})$ \cite{Bazavov:2009fk} implies that the contributions from the valence quarks are more important than those from the sea quarks. For $F_K/F_\pi$, the truncation of the expansion at NLO only involves the coupling constant $L_5$. With the MILC result for $L_5$, the NLO formula of $\chi$PT yields $F_K/F_\pi\NLO 1.16(4)$, again leaving little room for contributions of NNLO. Altogether, however, the increase in the value of the kaon decay constant generated by the quark masses is quite substantial: the MILC results for $F_K/F_\pi$ and $F_\pi/F_0$ imply $F_K/F_0=1.40(5)$. The NLO term of the chiral series dominates the correction also in this case: with the MILC values for $L_4$, $L_5$, I get $F_K/F_0\NLO 1.31(5)$, leaving 0.09(8) for the remainder from higher orders. This confirms that a truncation of the chiral perturbation series is meaningful, but  the accuracy is quite limited if terms of NNLO are neglected. 
\begin{figure}[thb]\label{fig:ms}
\centering
\includegraphics[width=8.3cm]{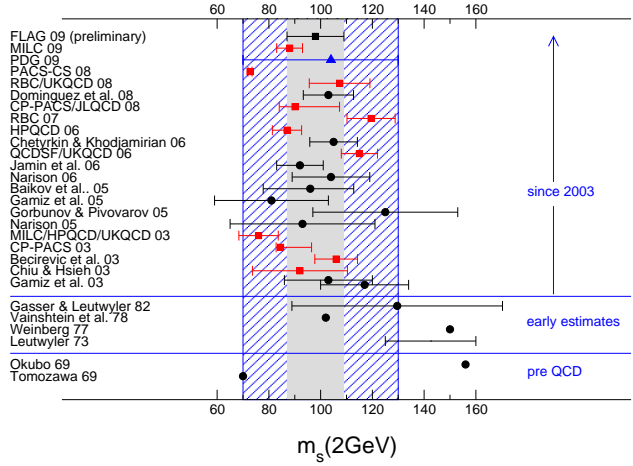}
\caption{Mass of the strange quark, in the $\overline{\mbox{MS}}$ scheme at running scale $\mu=2$ GeV. Except for some of the early estimates, disks indicate sum rule determinations, while squares represent lattice results.}
\end{figure}

Often, the perturbation series for the eigenvalues of the Hamiltonian is more rapidly convergent than the one for the matrix elements. This also occurs in $\chi$PT, where the corrections to the leading order formulae for the masses of the Nambu-Goldstone bosons are substantially smaller than those found for the decay constants. It turns out that, in the $\chi$PT representation for the masses, the NLO contributions from the valence quarks and from the Dirac determinant are of opposite sign and roughly cancel. Accordingly, the Gell-Mann-Oakes-Renner formula describes the dependence of $M_K^2$ on $m_s$ quite well, on the entire range between the chiral limit and the physical value of $m_s$: despite the fact that the mass of the strange quark is about 27 times heavier than the mean of $u$ and $d$, the departure from linearity is remarkably small. I will return to the convergence properties of the expansion in section \ref{sec:quark mass ratios}, in connection with the quark mass ratio $m_s/m_{ud}$. 

Note also that the numbers obtained at NLO are unambiguous only up to contributions of higher order.  Although this only matters beyond NLO, the convergence appears to favour the representation obtained by expressing the corrections in terms of the masses and decay constants of the Nambu-Goldstone bosons rather than the quark masses and couplings occurring in the effective Lagrangian. At least in part, this can be understood from the fact that the properties of $\chi$PT are governed by the infrared singularities of QCD. Since the position of these singularities is determined by the masses of the Nambu-Goldstone bosons, expressing the formulae in terms of these before truncating the series ensures that the singularities are sitting at the proper place, ab initio.  

The progress in the numerical simulation of QCD with light dynamical quarks is impressive, but Figure \ref{fig:ms} shows that the lattice determinations of $m_s$ do not yet yield a satisfactory picture  (for some of the lattice entries, only the statistical error is shown, because an estimate for the systematic error is lacking). One of the problems may arise from nonperturbative renormalization effects -- some of the collaborations still use perturbative renormalization. Also, since $m_s$ is often taken in the vicinity of the physical value, while $m_{ud}$ is substantially larger than in nature, the mass of the kaon is too large for the NLO formulae of $\chi$PT to yield a good basis for the extrapolation to the physical values. Within the present uncertainties, the lattice results confirm the values of $m_s$ found on the basis of QCD sum rules. It does not take much courage to predict that the progress being made with lattice simulations of light dynamical quarks will soon lead to a significantly more precise determination of $m_s$.
\section{Contributions from the electromagnetic interaction}\label{sec:electromagnetic interaction}
When comparing QCD calculations with experiment, radiative corrections need to be applied. In lattice simulations, where the QCD parameters are fixed in terms of the masses of some of the hadrons, the electromagnetic contributions to these masses must be accounted for.  

The decomposition of the sum ${\cal L}_{\mbox{\tiny QCD}}+{\cal L}_{\mbox{\tiny QED}}$ into two parts is not unique. The problem arises because the parameters $\alpha_s$, $m_u$, $\ldots$ need to be fixed in order to specify ${\cal L}_{\mbox{\tiny QCD}}$. These quantities depend on the renormalization scale and the dependence on the scale in the full theory differs from the one in QCD.  If ${\cal L}_{\mbox{\tiny QCD}}$ is specified by setting the values of $\alpha_s$ and of the quark masses equal to those relevant in the framework of the Standard Model, then the result depends on the scale at which the matching is done. The full theory is unambiguous, but the decomposition into two parts involves a convention. The value of the mass of the charged pion in QCD, for instance, is not a physical quantity, it is a matter of choice.

In pion physics, the standard convention used is such that, in the isospin limit, $M_\pi$ coincides with the physical mass of the charged pion and the radiative corrections that account for the effects due to the e.m.~interaction   are done accordingly (see, for instance, \cite{em corrections}). The reason for this choice is that the numerical values of scattering lengths and other quantities of physical interest are usually given in units of the physical value of $M_{\pi^+}$ -- working with a different value of the pion mass would lead to a rather clumsy framework. The convention is perfectly legitimate, but it corresponds to matching at a scale that is very different from the standard choice $\mu=2$ GeV, made when quoting quark mass values. In order to give results for the quark masses in the Standard Model at scale $\mu=2\,\mbox{GeV}$, on the basis of a calculation done within QCD, the two theories must be matched at this scale. In the following, I adopt this  convention.

The electromagnetic interaction plays a crucial role in determinations of the ratio $m_u/m_d$, because the isospin breaking effects generated by this interaction are comparable to those from $m_u\neq m_d$. In determinations of the ratio $m_s/m_{ud}$, the electromagnetic interaction is less important, but at the accuracy reached, it cannot be neglected. The reason is that, at leading order in the chiral expansion, this ratio is inversely proportional to $M_\pi^2$. Since the pion mass represents a small symmetry breaking effect, it is rather sensitive to the perturbations generated by QED. 

The information available about the e.m.~self energies of $\pi^0$ and $\pi^+$ is meager. The Cottingham formula \cite{Cottingham} represents these as an integral over electron scattering cross sections; elastic as well as inelastic reactions contribute. For the charged pion, the term due to elastic scattering, which involves the square of the e.m.~form factor, makes a substantial contribution. In the case of the $\pi^0$, this term is absent, because the form factor vanishes on account of charge conjugation invariance. Indeed, the contribution from the form factor to the self-energy of the $\pi^+$ roughly reproduces the observed mass difference between the two particles.\footnote{A reevaluation on the basis of the present knowledge would be of considerable interest.}
This does not imply, however, that the electromagnetic self-energy of the $\pi^0$ is negligibly small -- the size of the inelastic contributions is not reliably known. The low energy theorem of Das, Guralnik, Mathur, Low and Young \cite{Das et al 1967} ensures that, in the limit $m_u,m_d\rightarrow 0$, the e.m.~self-energy of the $\pi^+$ is given by an integral over the difference between the vector and axial spectral functions,\addtocounter{footnote}{-1}\footnotemark{} while the one of the $\pi^0$ vanishes, but the size of the corrections of order $e^2 M_\pi$ is difficult to estimate. A more accurate evaluation of the self-energy of the pion is required in order to improve the  quark mass determinations on the lattice. At the present level of accuracy, it does not matter whether this is done for the charged or for the neutral pion, because the uncertainty in the difference between charged and neutral pion masses in QCD is small compared to the one in the self-energies. In the absence of a good determination of the e.m.~self energy on the lattice, I think that it is advisable to use a conservative estimate, allowing inelastic processes to contribute at the level of 1 MeV. This limits the accuracy to which the ratio $m_s/m_{ud}$ can presently be determined to about $2\times (1\, \mbox{MeV}/135\,\mbox{MeV})\simeq 1.5\%$.
  
\section{Quark mass ratios}\label{sec:quark mass ratios}
\begin{figure}[thb]\centering
\includegraphics[width=9cm]{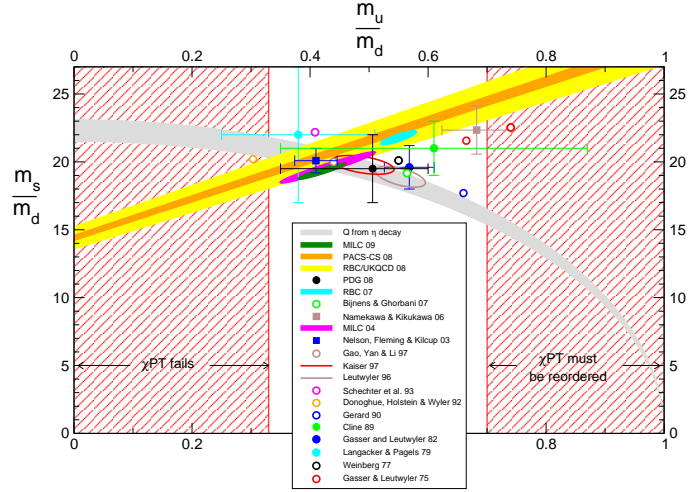} 
\caption{\label{fig:ms/md versus mu/md}
Ratios of the light quark masses}
\end{figure}

At tree level of the effective theory, the masses of the Nambu-Goldstone bosons obey the Gell-Mann-Oakes-Renner formulae $M_{\pi^+}^2\LO B_0(m_u+m_d)$, $M_{K^+}^2\LO B_0(m_u+m_s)$, $M_{K^0}^2\LO B_0(m_d+m_s)$. The constant $B_0$ is related to the quark condensate, but since $\chi$PT does not predict its size, there is no prediction for the size of the quark masses, either. Their ratios, on the other hand, can be expressed in terms of the meson masses. Correcting these for the e.m.~self-energies with leading order $\chi$PT ("Dashen theorem") and neglecting contributions of second order in the isospin breaking mass difference $m_u-m_d$, one arrives at Weinberg's relations of 1977  \cite{Weinberg:1977hb}:
\be\label{eq:Weinberg 1977}\frac{m_u}{m_d}\LO\frac{M_{K^+}^2 - M_{K^0}^2 + 2 M_{\pi^0}^2 - M_{\pi^+}^2}{M_{K^0}^2 - M_{K^+}^2 + M_{\pi^+}^2}=0.56\,,\hspace{1cm} \frac{m_s}{m_d}\LO \frac{M_{K^+}^2 + M_{K^0}^2  - M_{\pi^+}^2}{M_{K^0}^2 - M_{K^+}^2 + M_{\pi^+}^2}=20.2\,.\ee 
These relations are valid only to leading order of the chiral perturbation series.
In view of the fact that a {\it massless $u$-quark} would solve the strong CP-problem, many authors have considered this an attractive possibility and some have even claimed that this solution only calls for modest contributions from higher orders. That, however, is not the case \cite{mu not 0}. If $m_u$ were equal to zero, then the above relation for $m_u/m_d$ would predict $M_{K^0}-M_{K^+}=16.9\,\mbox{MeV}$, to be compared with the observed mass difference of 3.9 MeV. It should be evident that a framework that treats the higher order contributions as corrections cannot recover from such a failure. A very generous range for which a truncation of the chiral expansion is not a priori meaningless is $ 0.25<m_u/m_d<0.7$ (see Figure \ref{fig:ms/md versus mu/md}; the reordering of the series required above 0.7 is discussed in \cite{Stern}).  The conclusion to draw if $m_u$ were to vanish, would be that it is meaningless to truncate the chiral series at low orders, so that the success of the eightfold way cannot be understood within QCD. In effect, the proposal thus merely replaces one puzzle by the other. The work done on the lattice should close this chapter, confirming that a massless $u$-quark represents an interesting way not to understand this world: none of the lattice results is consistent with $m_u=0$. In particular, the MILC collaboration rules this solution out at 10 $\sigma$. Nature solves the strong CP problem differently.

$\chi$PT predicts the two quark mass ratios in terms of meson masses only at leading order. At NLO,  chiral symmetry implies only one parameter free relation between the quark masses and the meson masses of QCD \cite{GL SU3}:
\be Q^2\equiv\frac{m_s^2-m_{ud}^2}{m_d^2-m_u^2}\NLO \frac{{M_K}^2-M_\pi^2}{M_{K^0}^2-M_{K^+}^2}\cdot\frac{M_K^2}{M_\pi^2} \fs\ee
The values in (\ref{eq:Weinberg 1977}) imply $Q = 24.3$. The critical input here is the Dashen Theorem: the e.m.~self-energies are accounted for only at tree level of the effective theory. Alternatively, $Q$ can be extracted from the experimental information concerning the decay $\eta\rightarrow 3\pi$. Since the electromagnetic contributions to this transition are suppressed, this determination of $Q$ is less sensitive to the uncertainties therein. A comprehensive analysis of this decay  is under way \cite{Lanz}. The gray elliptic band in Figure \ref{fig:ms/md versus mu/md} corresponds to the range $Q=22.3\pm 0.8$, which in my opinion is a fair assessment of the current knowledge based on $\eta\rightarrow 3\pi$. The MILC results for the quark mass ratios \cite{Bazavov:2009fk} imply $Q=21.7\pm 1.1$ and are thus consistent with the above value, but those obtained by the RBC collaboration \cite{RBC}, which are based on a simulation with $N_f=2$, disagree with it, as they correspond to $Q=26.1\pm 1.2$. 

The position on the ellipse cannot be determined on the basis of phenomenology alone. The expansion in powers of $1/N_c$ does give a theoretical handle. Unfortunately, however, the bound I had obtained in that framework \cite{Leutwyler 1996} receives large corrections from higher orders \cite{Kaiser}. The experimental information about the width of the decays $\eta\rightarrow\gamma\gamma$ and $\eta'\rightarrow\gamma\gamma$ can be used to bring the $1/N_c$ expansion under better control. The resulting pattern for the masses and mixing angles of the pseudoscalar nonet implies $m_s/m_{ud}=26.6\pm 1.6$ \cite{Kaiser}, indicating that the corrections to the value $m_s/m_{ud}=25.9$ that follows from the leading order ratios (\ref{eq:Weinberg 1977}) are small. 

The lattice results for $m_s/m_{ud}$ are slightly higher. Adding the quoted errors\footnote{These do not account for the uncertainties in the electromagnetic self-energies discussed in section \ref{sec:electromagnetic interaction}.} in quadrature, the MILC result reads $m_s/m_{ud}=27.4(2)$  \cite{Bazavov:2009fk}. As a check on the convergence of the expansion in this case, it is instructive to evaluate the NLO formula,\footnote{The constant $F_\pi$ can be replaced by $F_0$ -- the operation merely affects the size of the NNLO corrections.}\be\label{eq:ms/mud NLO}\frac{m_s}{m_{ud}}\NLO\frac{2M_K^2}{M_\pi^2}
\left\{1-8\,\frac{M_K^2-M_\pi^2}{F_\pi^2}(2L_8-L_5) +\mu_\pi-\mu_\eta\right\}-1\co\ee 
where the chiral logarithms stand for $\mu_P=M_P^2 \;\ell n\, (M_P^2/\mu^2)/(32\pi^2 F_\pi^2)$. For the relevant combination of effective coupling constants at running scale $\mu=M_\eta$, the MILC collaboration quotes the value  $2 L_8 - L_5 = - 0.48 (8)(21)\times 10^{-3}\hspace{-0.1cm}.\hspace{0.1cm}$ This leads to $m_s/m_{ud}\NLO 28.1(0.5)(1.2)$. Although the uncertainty in the couplings still leaves room for contributions from higher orders, the NLO formula does represent a decent approximation. 
\section{Conclusions}
\noindent
\p $\chi$PT based on SU(2)$_L\times$SU(2)$_R$ has become a precision tool.\\
\p The lattice yields remarkably coherent and significant results for pion physics already now.\\
\p $m_u\neq 0$. Nature solves the strong CP problem differently.\\
\p The lattice results for $m_s$ are consistent with the values obtained from sum rules.\\
\p The extension to SU(3)$_L\times$SU(3)$_R$, i.e.~to the physics of the  strange quark, is making progress.\\
\p The lattice results indicate that,  at the physical values of the quark masses, the extension works,\\\rule{0.25cm}{0cm}  but configurations  with $M_K \geq 600$ MeV are beyond reach of NLO $\chi$PT.\\
\p For $M_\pi,M_K,M_\eta$, leading order $\chi$PT yields a decent approximation, but for some of the matrix\\ \rule{0.27cm}{0cm} elements, $F_K$ for instance, the corrections are large, 
NNLO contributions cannot be neglected.\\
\p Representations for many quantities of interest are available to that order \cite{Bijnens proceedings}.\\
\p The main problem at NNLO is that the knowledge of the relevant effective coupling constants\\\rule{0.25cm}{0cm} is still rudimentary, but the lattice determinations of the couplings are steadily improving.\\
\p Significant progress at the interface between lattice and effective field theory is ante portas.

\end{document}